\documentstyle[aps,prl,twocolumn,epsf]{revtex}
\begin{document}
\title{Nonlinear feedback for control of chaos}
\author{M. de Sousa Vieira\cite{email} and A. J. Lichtenberg} 
\address{Department of Electrical Engineering and Computer Sciences, University of California, Berkeley, CA 94720, USA.}
\maketitle
\begin{abstract}
We generalize a method of control of chaos which uses delayed feedback 
at the period of an unstable orbit to stabilize that orbit. The 
generalization consists of substituting some portion of the nonlinear
dynamical system with a delayed dynamics, rather than using a 
linear delay function for control. A further generalization, in which the 
control function retains memory of all previous periods, allows 
the region of the parameter space over which control can be achieved  
to be extended, but at the price of losing the ability to achieve 
superstability. Nonliner feedback results in a larger  
basin of attraction to the stabilized orbit than linear 
feedback. For a simple test mapping studied (the logistic map) the 
dimension of the system increases from one to two by introducing 
control. We show in the case involving memory, for a particular 
choice of the relationship between the control parameters, that the 
superstable orbit can be recovered without reducing the parameter 
space  that can be controlled. This particular solution, in addition 
to having the largest basin of attraction of the methods considered, 
retains the dimension of the uncontrolled system. 
\end{abstract} 
\narrowtext
\section{Introduction}\label{s1}

A number of methods have been proposed for feedback control of 
chaos\cite{huebler,ogy,pyragas,socolar,bielawski}.  
Two methods of control that stabilize an otherwise unstable periodic orbit 
have received considerable attention recently: 

(a) Ott, Grebogi  and Yorke (OGY)\cite{ogy} introduced a method 
which stabilizes unstable 
periodic orbits (UPO's) found in the chaotic regime via small feedback 
perturbations to an accessible parameter. 
The control perturbation is given when the orbit crosses a given 
Poincar\'e section, such that the trajectory will be  close to the 
stable manifold of the desired UPO. In this method, in the limit of zero 
noise, the orbit of the controlled system is identical to the 
UPO of 
the uncontrolled system and the feedback 
perturbation vanishes.  
A drawback for the OGY method is that 
it becomes difficult to apply for very 
fast systems, since it requires  computer analysis of the system 
at each crossing of the Poincar\'e section. Also,   
noise can result in occasional bursts where the trajectory wanders 
far from the controlled periodic orbit.   

(b) An alternative method of feedback stabilization of UPO's, 
introduced by 
Pyragas\cite{pyragas},
consists of a continuous linear feedback 
applied  
at each computational time step. As in the OGY case, in  this 
method the controlled orbit coincides with the UPO of 
the uncontrolled system 
and the feedback vanishes, for zero noise, when control is achieved. 
The feedback procedure can be applied without 
knowing a priori the location of the periodic orbit, for a version in which 
the feedback term contains a delayed variable, in which the delay 
corresponds to the period of the UPO.  Moreover, it is  
expected that it can be used for fast systems, since 
no parameters are changed on a fast time-scale, 
and the method does not require a 
computer analysis of the system.   
For some systems, 
the method is robust even in the presence of  
considerable noise\cite{tamasevicius}. 
A disadvantage of Pyragas' method is that it achieves control only over  
a limited range of the parameter space, i.e., a given orbit will 
become eventually unstable in the controlled system as the parameters 
are varied more deeply into the chaotic regime.  
The use of delayed feedback also increases the dimensionality of the system. 
Socolar {\sl et. al.}\cite{socolar} extended the Pyragas method 
to include, in the control term,  memory of all the previous 
states of the system, and were thereby able to increase the 
region of the parameter space where control can be achieved.   
The dimensionality of the system 
also increased in this method, to the same extent as in the method 
of Pyragas. 

The  desirable properties of a control system depend on the application. 
Here we consider a system with a fixed period UPO which we are 
attempting to control. The parameters that control the coordinates of the 
UPO may be slowly varying compared to the UPO period. The system can 
be considered subject to noise, which may take the system coordinates 
away from the periodic orbit. In this general situation we may consider 
the following properties of the control system as desirable: (1) In the 
neighborhood of the periodic orbit (assumed stably controlled)  
the actual  orbit returns  optimally fast to the periodic 
orbit when perturbed away from it (in the limiting case the orbit is 
superstable). (2) A slow drift in parameters can be tracked by the 
control over the largest possible parameter space. (3) In the larger 
space we (nonlinearly) want the stability to be maintained over 
the widest range of uncertainty in either the system parameters or the 
dynamical variables (basin of attraction) caused by noise. 
(4) In some average sense, we wish to minimize the 
time required to return to the 
desired solution from the basin 
of attraction (in the nonlinear regime).  

In the next section we examine these criteria for control of a 
fixed point of a mapping corresponding to a periodic solution 
of a continuous system, with a known period. We use the 
well studied logistic map as the test bed for the study. 
Here we introduce modifications to the control methods  of 
Pyragas and Socolar, using a nonlinear
function in the feedback term. 
As in the OGY, Pyragas and Socolar methods, 
the stabilized orbit 
is identical to the UPO of the uncontrolled system, 
and when control is achieved the magnitude of the feedback 
term vanishes in the absence of noise.  
First we compare the linear control of Pyragas to an 
analogous nonlinear control. Next, we compare the case in which 
memory is introduced into the mapping parameters with case in 
which  
memory is introduced into the nonlinear control. 
In Section~\ref{s3} we introduce 
a special case of nonlinear control with memory, that reduces 
to a remarkable simple form, whose properties are generally better 
than the other cases examined. We show that this latter 
control can be used efficiently to control other periodic 
points and higher dimensional mappings.   

\section{Comparison of continuous control methods}\label{s2}

We start by 
describing Pyragas' method\cite{pyragas}. 
He considered a dynamical system that is governed by ordinary differential 
equations, which are in principle unknown. However, some scalar variable $y$ 
can be measured as a system output, and  the system  also has an input available 
for an external force $f$. These assumptions can be met by the following 
model,
\begin{equation}
{{d y}\over {dt}}= P(y,{\bf x}) + f(t),\ \ \ {{d {\bf x}}\over{dt}}={\bf Q}(y,\bf x), 
\label{eq1}
\end{equation}
where {\bf x} describes the remaining variables of the dynamical system which 
are not available for observation  or not of interest. The forcing term 
disturbs only the variable $y$, and it is assumed that the  
system may be in the chaotic regime when  the forcing term $f(t)$ is zero. 

Pyragas studied two types of forcing. In the first method one determines 
the UPO $y_i$ of the chaotic attractor from $y(t)$, 
following well 
known algorithms\cite{lathrop}. Then one designs an oscillator that has 
an orbit equal to that of $y_i$. The forcing 
term is given in this case by 
$f(t)=K[y_i(t)-y(t)]$, where $K$ is an empirically adjustable 
weight of the perturbation.  
In the other type of forcing considered by Pyragas, the forcing term   
contains a delayed term of the variable $y$, namely,  
$f(t)=K[y(t-\tau)-y(t)]$, where $\tau $ is the delay time. If the delay 
time coincides with period of the $i$-th UPO  then the perturbation $f(t)$ 
vanishes and $y(t)$ will coincide with UPO, as in the first case. 
However, in this 
last case, one does not need to know the     
UPO, just its period, nor is it necessary to design an external oscillator. 
[Although Pyragas described his method for the 
situation in which one knows only a time series,  
in all the cases he studied the equations that described the system 
were known.]  
Here we are concerned with the second method, i.e., the delayed
feedback case. 

Applying Eq.~(\ref{eq1}) for the stabilization of a period-one orbit in 
a one-dimensional mapping $F(x_n)$, 
we have
\begin{equation}
x_{n+1}=F(x_n)+K[x_{n}-x_{n-1}].
\label{eq3}
\end{equation}
The controlled system has dimensionality equal to two instead of one for 
the unperturbed system.   
The eigenvalues of Eq.~(\ref{eq3}) are given by expanding it about the 
equilibrium $x_{n+1} = x_n$ to obtain, 
\begin{equation}
\lambda_{1,2}={{F'+K \pm \sqrt{[F'+K]^2-4K}}\over {2}}, 
\label{eq4}
\end{equation}
where $F' \equiv F'(x_f)$ is the derivative of $F$ with respect to $x_n$ at 
the fixed point $x_f$. 

We illustrate this method of control using the logistic map, 
\begin{equation}
x_{n+1}=F(x_n)=4ax_n(1-x_n). 
\label{eq2}
\end{equation}
This map presents a sequence of period doubling bifurcations as $a$ 
increases and enters into chaos at $a \approx 0.8925$. The period-one orbit is 
stable from $a=0$ to $a=3/4$.  
The fixed point $x_f$ for the period-one orbit is zero for 
$0 \le a <1/4$ and $x_f=1-1/4a$ for $1/4 < a <3/4$. 
If $a<0$ or $a>1$, the attractor is unbounded, diverging 
to infinity. The attractor will also diverge if the initial 
condition $x_0$  is not 
in the interval [0, 1].  
The period-one orbit loses stability when one of the eigenvalues has 
modulus larger than one. For the logistic map, for $K<1$,  
an eigenvalue crosses -1, causing the 
appearance of a pitchfork bifurcation. 
When this occurs, Eq.~(\ref{eq4}) gives $F'(x_f)=-1-2K$. Since,  
for the logistic map $F'(x_f)=2-4a$, the bifurcation point $a^*$ is   
\begin{equation}
a^* ={{3+2K}\over{4}}.  
\label{eq40}
\end{equation}
A Hopf bifurcation occurs at $K=1$, where 
the eigenvalues cross the unit circle with imaginary values.  
Beyond this value of $K$ there is no stable solution, so
the maximum value of $a$ where control can be achieved 
is given by $a^*=1.25$.  

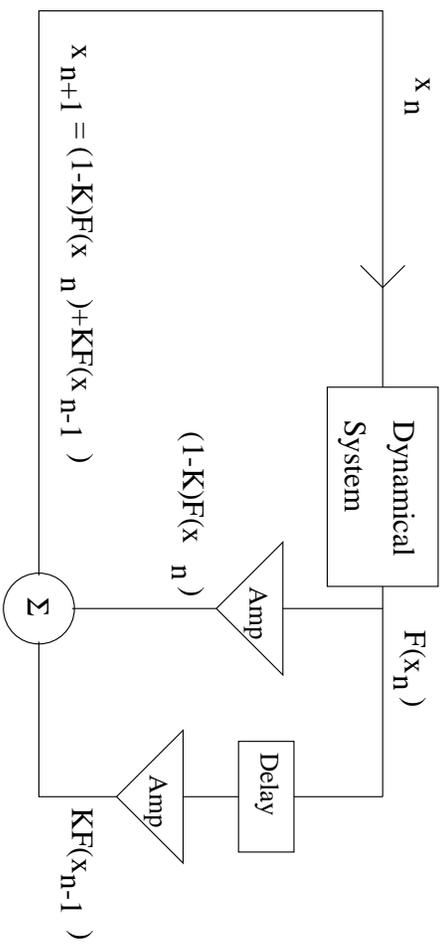
\begin{figure}
\caption[f1]{
Block diagram for the nonlinear feedback control of chaos without memory.  
}
\label{f1}
\end{figure}

We compare these results with the use of a  nonlinear rather than a linear
function in the feedback term.
Thus, the forcing term
to stabilize a periodic orbit is given by
is a nonlinear function $G(x_n,x_{n-1})$. 
Obviously, many choices can be made for $G$, with the constraint that
$G=0$ in the desired UPO.
Perhaps the simplest construction is $G=-K[F(x_n)-F(x_{(n-1)}]$, with 
$K>0$,  since 
with this feedback one does not need to know the equation $F(x_n)$ that 
governs the system. We show how this control could be applied in the block 
diagram displayed in Fig.~\ref{f1}.  This 
feedback also gave a better performance than 
other nonlinear functions, as discussed below.  

For the period-one orbit, our controlled system can be written as 
\begin{equation}
x_{n+1}=F(x_n) - K[F(x_{n})-F(x_{n-1})].
\label{eq5}
\end{equation}
The eigenvalues for this equation are
\begin{equation}
\lambda_{1,2}={{(1-K)F'\pm\sqrt{[(1-K)F']^2 + 4KF'}}\over
 {2}}.
\label{eq52}
\end{equation}
When a pitchfork bifurcation occurs,  for $K$  not very large,
the most negative eigenvalue is equal $-1$. From Eq.~(\ref{eq52}) we 
obtain  
$F'(x_f)=-1/(1-2K)$.
For the logistic map 
this 
gives
\begin{equation}
a^*={{3-4K}\over{4(1-2K)}}.
\label{eq55}
\end{equation}
From Eq.~(\ref{eq55}) we see that if $K<0$,   
then $a^*<3/4$. 
For $K > 1/3$, the period-one loses its stability not via a period doubling 
bifurcation, but via a Hopf bifurcation.
For this case, we obtain
\begin{equation}
a^*={{1+2K} \over {4K}}.
\label{eq9}
\end{equation}
With Eqs.~(\ref{eq55}) and ~(\ref{eq9}) we find that the maximum value for $a$  
where control can be achieved with this method is also $a^*=1.25$, which  
occurs at $K=1/3$..  
We compare $a^*$ as a function of $K$ for the nonlinear control  
in Fig.~\ref{f2} (solid line) to that using the linear  control 
of the Pyragas' method (dashed line). 
In Fig.~\ref{f3} we compare the values of $\log_2 |\lambda|$, with $\lambda$  
being the least stable eigenvalue, for nonlinear control 
(solid line) with linear feedback (dashed line) and with no control 
(short-dashed line). We note that the 
transient time (in the linear regime), which is proportional to 
$1/|\log_2 |\lambda||$, is smaller in the nonlinear control than in 
Pyragas's method. Also,   
superstable orbit, 
$\log _2 |\lambda|= - \infty$, 
is preserved with the nonlinear control.  
[The reader might well ask why the parameters linearized around the fixed 
point are not the same for linear and nonlinear control. As pointed out 
by a referee, the use of a linear (in $x$) control parameter with  
$K \to K(4a-2)$ brings the results into coincidence. However, 
a dependence of the control on $a$ introduces a new complexity into the 
feedback, which we consider below].  We note in Fig.~\ref{f2} that 
the largest range of stable $a$, $a<a^*$, occurs at $K=1$ for the linear 
control. This has significant disadvantages when we consider the nonlinear 
phase space, as we now show. 
   
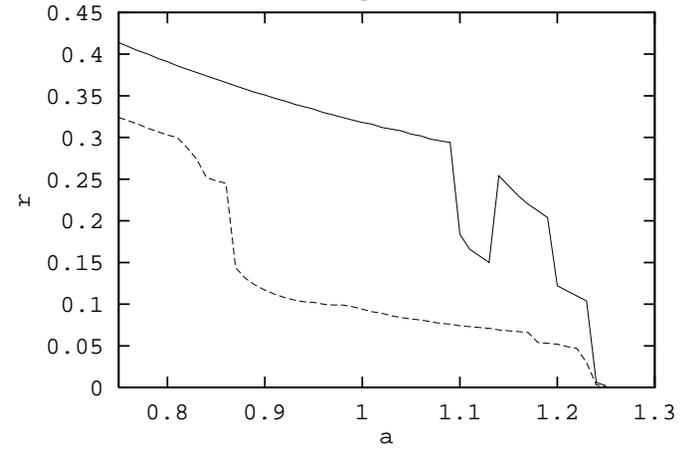
\begin{figure}
\caption[f4]{
(a) Basin of attraction and noise circle for (a) the nonlinear 
control without memory 
for $a=1$ and $K=1/3$  
and for (b) Pyragas' linear control for $a=1$ and $K=0.75$.   
}
\label{f4}
\end{figure}

We numerically determined the basin of attraction by constructing a 
grid of initial conditions in the $x_{n-1}$, $x_n$ space and determining 
which are attracted to the fixed point. The effect of noise on the stability 
is qualitatively examined by constructing a noise circle around the 
fixed point, which just touches the basin boundary, and finding the 
radius $r$ of the circle. We illustrate these properties in Fig.~\ref{f4}(a) 
for the nonlinear control and in Fig.~\ref{f4}(b) for the linear control. 
In Fig.~\ref{f5} we show how $r$ varies with $a$ for the nonlinear 
control at $K=1/3$ (solid line) and for the linear control along 
the dotted line of Fig.~\ref{f2} (dashed line). We see that the 
nonliner control is more robust in the presence of noise. 

Although the control procedure we have used is straightforward to 
implement, the reader might well ask if a different form of nonlinearity 
might be better. To investigate this possibility we studied the 
quadratic map 
\begin{equation}
x_{n+1}=1-ax_n^2.
\label{eq16}
\end{equation}
We found that the control term that is most robust in the presence of  
noise and has the smallest transient is given by 
$F=Ka(|x_{n}|^z-|x_{n-1}|^z)$ with $z=2$. 

Another method of control of chaos that follows Pyragas' ideas was introduced
by Bielawski {\sl et.~al.}\cite{bielawski}. In this method  the forcing is given
 to an
accessible {\sl parameter} of the
system, instead of adding a feedback term to the equation. The controlled logistic map in
 this case is given by
\begin{equation}
x_{n+1}=4(a+\epsilon_n)x_n(1-x_n),
\label{eq10}
\end{equation}
with $\epsilon_n={{K}\over{4}}(x_n-x_{n-1})$. The method has been 
generalized by Socolar {\sl et. al.}\cite{socolar}, with the controlled 
logistic map given by Eq.~(\ref{eq10}), but with 
\begin{equation}
\epsilon_n={{K}\over{4}}(x_n-x_{n-1})+R\epsilon_{n-1}, 
\label{eq101}
\end{equation}
where 
$R<1$.  The case $R=0$ reduces to the Bielawski control. 
Socolar {\sl et. al.}\cite{socolar} have shown that this 
form  of the control parameter is equivalent to including 
memory of all the past 
states of the system. The dimensionality of the new map is also two
with the variables $x_n$ and $\epsilon _n$. 
The eigenvalues are  
\begin{equation}
\lambda_{1,2}={{F'+R+\gamma \pm\sqrt{[F'+R+\gamma]^2-4[F'R+\gamma]}}\over {2}},
\label{eq11}
\end{equation}
where 
$\gamma=K(4a-1)/(4a)^2$.
A pitchfork bifurcation occurs at 
\begin{equation}
K=8a^{*2}(R+1)(4a^*-3)/(4a^*-1). 
\label{eq20}
\end{equation}
A Hopf bifurcation occurs at  
\begin{equation}
K=16a^{*2}[1+ 2R(2a^*-1)]/(4a^*-1). 
\label{eq21}
\end{equation}
The stability boundaries $a^*(K,R)$ are given by Eqs.~(\ref{eq20}) 
and ~(\ref{eq21}). By varying $R$ one finds that, 
in the absence of noise, control can be achieved for arbitrary 
large values of $a$. However, the width of the window in $a$ 
where control can be achieved decreases as $a$ increases. 

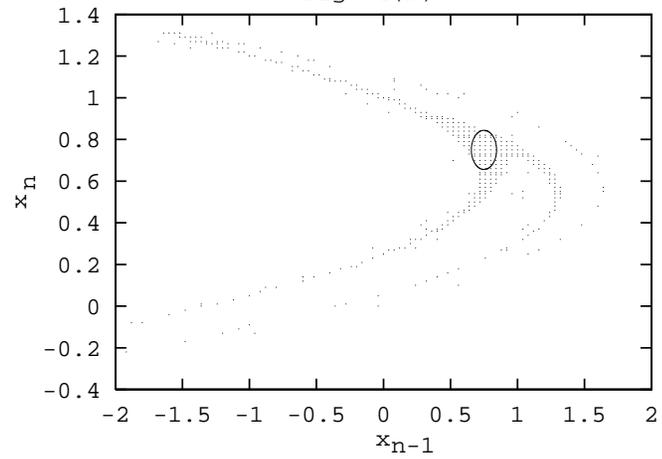
\begin{figure}
\caption[f6]{
Boundary of stability for the period-one orbit using  our 
nonlinear control 
method with 
memory (solid line) and Socolar's control method (dashed line), with $R=0.5$  
in both cases.  
}
\label{f6}
\end{figure}

Memory can also be included in the form of nonlinear control by 
a generalization of Eq.~(\ref{eq5}), 
\begin{equation}
x_{n+1}=F(x_n) +\epsilon _n
\label{eq60}
\end{equation}
\begin{equation}
\epsilon_{n+1}=-K[F(x_{n+1})-F(x_n)]+R\epsilon _n,
\label{eq61}
\end{equation}
When $R=0$ this control reduces to the case of nonlinear control studied 
above. 
The eigenvalues are 
\begin{equation}
\lambda_{1,2}={{(1-K)F'+R \pm\sqrt{[(1-K)F'+R]^2+4(K-R)F'}}\over {2}}, 
\label{eq71}
\end{equation}
and the stability boundaries for the logistic map are obtained from 
\begin{equation}
a^*={{3(1+R)-4K}\over{4(1+R)-8K}}  
\label{eq72}
\end{equation}
for a pitchfork bifurcation, and
\begin{equation}
a^*={{1}\over{2}} + {{1}\over{4(K-R)}}  
\label{eq733}
\end{equation}
for a Hopf bifurcation. The maximum value of $a$ where control 
can be achieved in this method is $a^*={{5-R}\over{4(1-R)}}$, which 
occurs at $K={{(R+1)^2}\over{R+3}}$. 
In Fig.~\ref{f6} the stability boundaries are shown for the 
mappings given by Eqs.~(\ref{eq60}) and ~(\ref{eq61}) (solid line) and 
Eqs.~(\ref{eq10}) and ~(\ref{eq101}) (dashed line), with $R=0.5$. 
For both mappings, the addition of a memory term (finite $R$) 
extends the region in parameter space that $a$ can be tracked. 
There is a distinct difference in the results of Fig.~\ref{f6} 
for the two methods of control. For the Socolar method the range 
of $a$ that can be stabilized becomes small, as $K$ tracks $a$ to 
large values. In contrast, the additive nonlinear control picks out 
a value of $K$ for which the mapping can be controlled for all 
values of $a$ up to a maximum, and thus is not sensitive to 
parameter drift or uncertainty. The range of $a$ that can be 
controlled at the fixed point increases without bound, as 
$R \to 1$. 
 
There is a price to pay for having a finite $R$ when the 
variable is subject to noise. For example, at $R=0.5$, $a=1$,  
$K=0.6428$ for the additive nonlinear control and $K=1.8333$ for the 
parameter control, which correspond, respectively, to the 
diamond and cross symbols 
shown in Fig.~\ref{f6}, we numerically find the 
basin of attraction for the two cases.  
The result for additive nonlinear control is shown in  Fig.~\ref{f7}(a)
and for control in the parameter in Fig.~\ref{f7}(b). 
In both cases the stabilization region has been decreased from 
that without memory, but much more so with parameter control
for which the basin appears to be fractal. 
The calculation of the noise radius here is somewhat subtle, 
since $\epsilon $ does not have a clear physical meaning.  If 
we add a noise term $\delta x$ to the right hand side of Eqs.~(\ref{eq10}) 
and ~(\ref{eq101}) we note that, when the system is at the 
fixed point, the variation in $x_{n+1}$ will still be $\delta x$. 
However, the variation in $\epsilon_{n+1}$ will be 
$(1+{{K}\over{4}})\delta x$. If the same procedure is applied to 
Eqs.~(\ref{eq60}) and ~(\ref{eq61}) we find that 
the variation in $\epsilon_{n+1}$ will now be  
$(1-KF'(x_f))\delta x$. Therefore the noise $\delta x$ is amplified in the 
$\epsilon $ variable in both cases, with distinct multiplicative factors.  
To compensate for this, we contract the 
$\epsilon $ coordinate by the respective factor before calculating 
the noise radius in the basin of attraction of the additive 
nonlinear control and the parameter control. Our results 
are shown in Fig.~\ref{f8} for $R=0.5$, for  
nonlinear additive control (solid line), calculated at $K=0.6428$,   
and for parameter control (dashed line), calculated at the 
median line of the stability boundary shown in Fig.~\ref{f6}.
We find that the nonlinear additive control is  more 
robust to noise than the parameter control. However, for the 
same value of $a$ the nonlinear control with $R>0$  is less robust 
than the case with $R=0$, which is shown 
in Fig.~\ref{f5}.     
We must also consider the effect
of varying $R$. We do this only for the case of the additive nonlinear
control in the next section.

\section{An optimal control function }\label{s3}

Although, the nonlinear control, with a memory factor $R$, has 
a number of desirable properties, there is the drawback that   
for a given $R$ at the value of $K$ for which $a^*$ obtains
its maximum value, there is no superstable orbit (this also occurs 
in the parameter control). 
Since operation at a superstable orbit is very desirable
from the perspective of return to the fixed point solution in the 
presence of noise, we look for a relation between $K$ and $R$ for 
which a superstable orbit is recovered by setting $\lambda =0$ in 
Eq.~(\ref{eq71}). We find two solutions,  
\begin{equation}
R=0, \ \ \ F'(x_f)=0, 
\label{eq73}
\end{equation}
that, for the logistic map, corresponds to $a=0.5$, which is the 
solution without memory,  
and 
\begin{equation}
R=K, \ \ \ K={{F'(x_f)}\over{F'(x_f)-1}},   
\label{eq744}
\end{equation}
which, for the logistic map, corresponds to $ K={{4a-2}\over{4a-1}}$. 
We call the second solution, Eq.~(\ref{eq744}), an optimized control 
function, as it allows operation with superstability, i.e., with 
maximum control at the fixed point, for $0 \le a  < \infty$. 
We find that this solution has other desirable properties, such 
as a large basin of attraction, and, remarkably, {\sl reduces the 
phase space to a single degree of freedom}. 

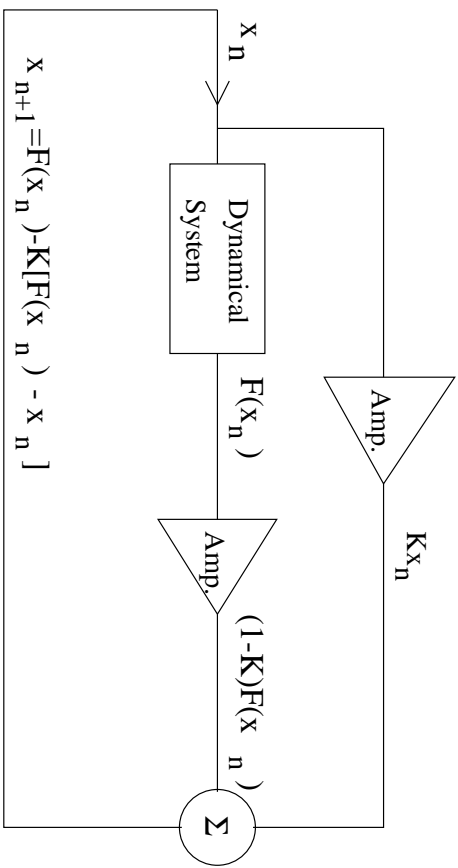
\begin{figure}
\caption[f9]{
Block diagram for our optimal control method.
}
\label{f9}
\end{figure}

Substituting $R=K$ into  Eq.~(\ref{eq61}), and  
eliminating 
$\epsilon _n$  
in favor of $x_{n+1}$ by using Eq.~(\ref{eq60}), we  
obtain 
$\epsilon_{n+1}=-K[F(x_{n+1})-x_{n+1}]$.
Dropping the index by 1, and substituting for $\epsilon _n$ in 
Eq.~(\ref{eq60}), 
we obtain a remarkably simple form for the mapping equation
\begin{equation}
x_{n+1}=F(x_n)-K[F(x_n)-x_n], 
\label{eq76}
\end{equation}
which is valid for control of the period-one orbit in one-dimensional maps. 
At the fixed point, the magnitude of the feedback term vanishes, as in the 
other methods studied here.
A block diagram of the optimal control scheme is 
shown in Fig.~\ref{f9}. One expects 
this method to be easy to implement in experiments, since
the control term contains only amplified versions of the input
and output of the dynamical system and one does not need to know 
$F$ to apply the control.
We note that for the particular case of a mapping, the period-one orbit 
is a fixed point. Thus the variable itself can be thought of 
as a delayed signal at the fundamental period of the updated variable. 
This property allows 
us to use a feedback signal with the same index as the mapping 
function itself. The delayed feedback is seen explicitly 
for control of 
differential equations, as discussed below.  
The eigenvalue 
for Eq.~(\ref{eq76}) is given by
\begin{equation}
\lambda=(1-K)F'(x_f)+K, 
\label{eq77}
\end{equation}
This map loses stability 
via a pitchfork bifurcation, where $\lambda=-1$. Consequently, the 
bifurcation point for the controlled logistic map is at
\begin{equation}
a^* ={{3-K}\over {4(1-K)}}.  
\label{eq78}
\end{equation}
Thus, we see that the 
parameter region where the 
period-one is stable increases as $K$ increases and tends to 
infinity as $K$ tends to one. 

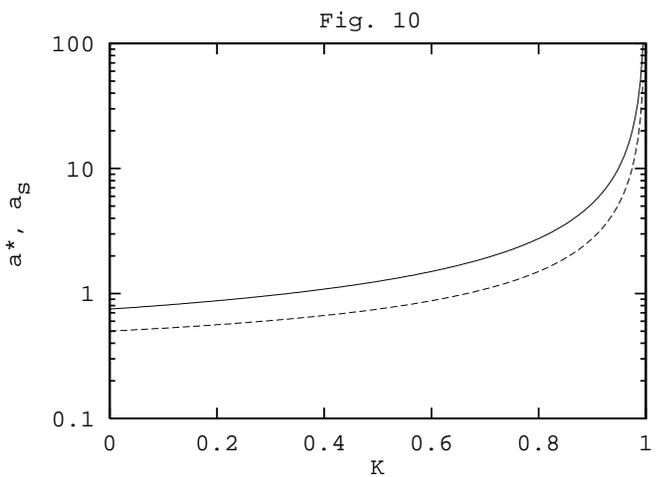
\begin{figure}
\caption[f10]{
$a^*$ and $a_s$ for the logistic map using the optimal control method.   
}
\label{f10}
\end{figure}

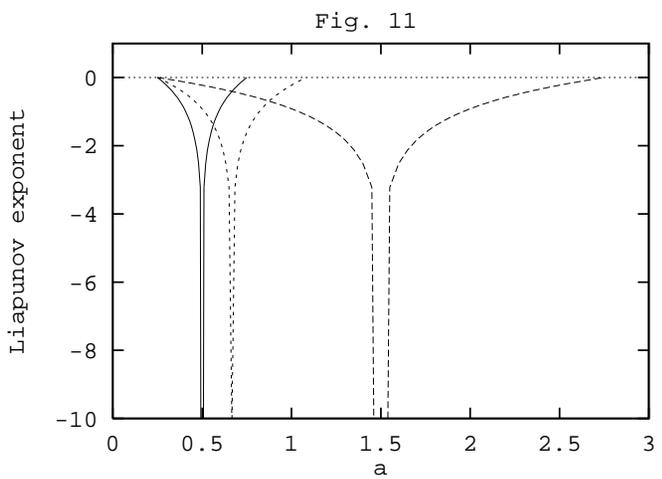
\begin{figure}
\caption[f11]{
Liapunov exponent, $\log _2 |\lambda |$, for $K=0$ (solid line), $K=0.4$ (dashed line), and $K=0.8$ 
(long-dashed line) for the logistic map using 
the optimal control method. 
}
\label{f11}
\end{figure}

The superstable orbit, $\lambda =0$, is obtained at 
\begin{equation}
a_s ={{2-K}\over {4(1-K)}},
\label{eq79}
\end{equation}
where the subscript `$s$' denotes superstable orbit. Also here 
$a_s$ increases with $K$ and goes to infinity as $K$ tends to 
one.  
In Fig.~\ref{f10} we show, as a function of $K$, 
the values $a^*$ where the 
period-one bifurcates (solid line) and the values $a_s$ of the 
superstable orbit (dashed line).   
We plot in Fig.~\ref{f11} the Liapunov 
exponent,  $\log _2|\lambda|$,  as a function of $a$,  for  
$K=0, 0.4, 0.8$. One can see from this figure that increasing $K$  
increases the range of the parameter $a$ around the superstable orbit 
for which a given transient time can be achieved.  

We now calculate the basin of attraction of the controlled UPO. Since 
our controlled map is one-dimensional this can be found
easily. 
Using Eq.~(\ref{eq76}) with $F$ given by Eq.~(\ref{eq2})   
the convergence to the UPO will be 
attained when  $0 \le x_0 \le 1+ {{K}\over{4a(1-K)}}$. 
Substituting for $a$ at the superstable orbit from Eq.~(\ref{eq79}) 
we obtain 
\begin{equation}
0 \le x_0 \le 1+ {{K}\over{2-K}}. 
\label{eq810}
\end{equation}
The basin of attraction increases with $K$, extending  
from 0 to 1 at $K=0$ to 0 to 2 at $K=1$. Since the fixed 
point is at $x_f=1-1/(4a)$, the noise radius around the fixed 
point is 
\begin{equation}
r =\min \left[ 1 - {{1}\over{4a}},\  {{K}\over {4a(1-K)}}+{{1}\over{4a}}\right], 
\label{eq80}
\end{equation}
which, at the superstable orbit, gives 
\begin{equation}
r_s={{1}\over{2-K}}, 
\label{eq81}
\end{equation}
such that  $r_s$ varies from 0.5 to 1 as $K$ varies from 0 to 1. 
Comparing Eq.~(\ref{eq81}) to our previous control parameters we 
see that this optimized control maintains stability better in the 
presence of noise. 
 
Although the superstable orbit is 
maintained, it is not clear what happens to the time constant for 
return to the periodic orbit as, $K \to 1$, for   
initial conditions that are started 
far away from
the fixed point.
To study the effect of  $K$ on the
nonlinear transient we do the following:
we start the system with
1000 different initial conditions, uniformly
distributed in the interval $[0,1+{{K}\over{4a(1-K)}})$.
Then we verify how many
iterations on average are necessary to bring the orbit
within a radius of $10^{-4}$ around the fixed point. The
result of the nonlinear transient as a function of
$K$ for $a= a_s$ is
shown in Fig.~\ref{f12}. 
It increases slightly as $K$ increases  
and goes to infinity at $K=1$, 
where Eq.~(\ref{eq76}) is marginally stable.
 
In a more general way, the nonlinear control with memory for 
stabilization of an UPO in a one-dimensional map with period $m$ is  
is given by
\begin{eqnarray}
x_{n+m}&=&F^m(x_n)+\epsilon_n\\
 \label{eq900}
\epsilon_{n+m}&=&- K[F^m(x_{n+m})-F^m(x_{n})]+R\epsilon_n. 
 \label{eq901}
\end{eqnarray}
For the case in which $K=R$ this control reduces to 
the optimized version 
\begin{equation}
x_{n+m}=F^m(x_n)-K[F^m(x_{n})-x_{n}] 
\label{eq811}
\end{equation}
Also higher periodic orbits the dimensionality of the controlled map is 
still one. 
The fixed points of the iterated map are identical to
the fixed points of the uncontrolled equation.
We have applied the optimized control for a period-two orbit ($m=2$) 
of the logistic map, in which the fixed points are given by 
$x_f=[4a+1\pm \sqrt{(4a-3)(4a+1)}]/8a$.
The eigenvalue for the period-two orbit can be easily
calculated and one finds that a pitchfork bifurcation from
period-two to period-four will
occur when $a^*={{1}\over{4}}[1+\sqrt{5+(1+K)/(1-K)}]$.
The superstable orbit is at $a_s={{1}\over{4}}[1+\sqrt{5+K/(1-K)}]$.
The value of $a$ where the bifurcation from period-one to
period-two occurs is at $a=0.75$, which is the same value found in the
uncontrolled map. Consequently, the region of the parameter
space where control can be achieved in the period-two orbit
also grows with increasing $K$, and goes to infinity as
$K$ tends to one. We 
note that the period-two orbit is also controllable by the methods 
considered in section~\ref{s2}, but are also ``non-optimal" in the 
sense that we have discussed. 

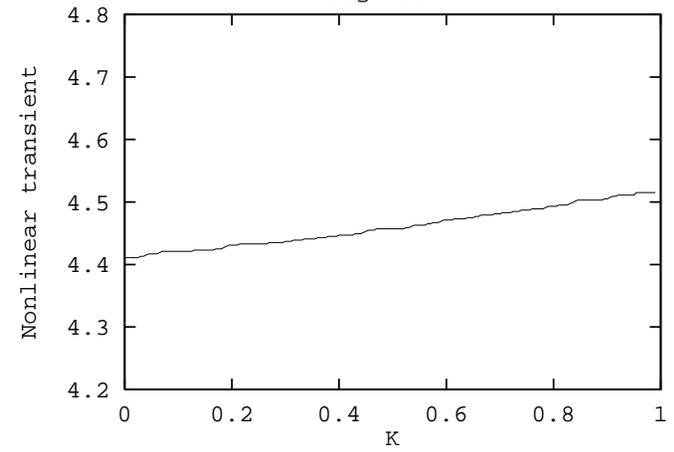
\begin{figure}
\caption[f13]{
The largest Liapunov exponent, $\log _2 |\lambda |$,  for the H\'enon map without 
feedback (label `Henon'), with 
feedback in the $x$ and $y$ variables (label `xy'), with feedback 
only in the $x$ variable (label `x'), and with feedback only in 
the $y$ variable (label `y'). 
}
\label{f13}
\end{figure}

Although the results of our study of controlling a simple 
one-dimensional mapping are suggestive of general underlying 
principles, they are not generic.  
A generalization of our ``optimized"  control scheme for a period-one orbit 
can be 
expressed in the following form
\begin{equation}
{\bf u}_{n+1}= {\bf P}({\bf u}_n ,{\bf v}_n) + {\bf f},\ \ \ {\bf v}_{n+1}={\bf Q}({\bf u}_n, {\bf v}_n),
\label{eq111}
\end{equation}
where ${\bf u}$ is a vector of the variables that are available for 
observation and  {\bf v} describes the remaining variables of the dynamical
system which
are not available or not of interest. The control term 
operates only on the ${\bf u}$ vector and is given by 
${\bf f}=K[{\bf u}_n - {\bf P}({\bf u}_n ,{\bf v}_n)]$.
We apply this more general form to a higher-dimensional mapping. For 
a specific example, we study the 
H\'enon map, which is given by 
\begin{equation}
x_{n+1}=1+y_n-ax_n^2, \ \ \
y_{n+1}=bx_n.
\label{eq161}
\end{equation}
In this map, for $b=0.3$ (which is the case we consider here),
the period one orbit is stable in the interval
$-0.1225 \alt a \alt 0.3671$. The system  enters into chaos  when $a \agt
 1.059$, and
the orbit becomes unbounded for $a \agt 1.428$.
For this map we can use three types of control: in both variables,
only in the $x$ variable or only in the $y$ variable.  For the
first type of control we have
\begin{eqnarray}
x_{n+1}&=&1+y_n-ax_n^2-K[1+y_n-ax_n^2-x_n],\\
y_{n+1}&=&bx_n-K[bx_n -y_n].
 \label{eq162}
\end{eqnarray}
For the second type of control the equations are
\begin{eqnarray}
x_{n+1}&=&1+y_n-ax_n^2-K[1+y_n-ax_n^2-x_n],\\ 
y_{n+1}&=&bx_n.
\label{eq163}
\end{eqnarray}
The third type of control gives  
\begin{eqnarray}
x_{n+1}&=&1+y_n-ax_n^2,\\  
y_{n+1}&=&bx_n-K[bx_n -y_n].
\label{eq164}
\end{eqnarray}
In all these cases the fixed points are the same as in the uncontrolled
H\'enon map, and the feedback term vanishes when control
is achieved.
We have found that  the control does not change the lower boundary of the
region of stability of the period-one orbit. However, the
upper boundary increases as $K$ increases for the three types of forcing.
For example, for $K=0.4$ the period-one orbit bifurcates
at $a\approx 1.98, 1.74, 0.49$ for the
the first, second and third methods, respectively.
Thus, different types of control have different regions for which  
stabilization is possible. 
For the H\'enon map the largest region of  
control occurs when  the  $x$ and $y$ variables are controlled simultaneously. 
The largest Liapunov exponent for the uncontrolled and for the
controlled H\'enon map is shown in Fig.~\ref{f13}, also for
$K=0.4$. As we see,
no superstable orbit exists for any $a$,  
for the period-one orbit in the uncontrolled equation with $b=0.3$.
The feedback terms we use 
to expand the region of stability modifies the location of the 
most stable orbit but do not create a superstable orbit when one 
does not exist for any value of $a$ in the uncontrolled equation. 

\section{Conclusions and Discussion}\label{s4}

We have generalized a method introduced by Pyragas\cite{pyragas}, used
to control an otherwise unstable periodic orbit, as applied to mappings.
The method consists of feeding back a delayed signal with the delay 
equal to the period to be controlled, done in such a manner that the 
position of the stabilized orbit in the phase space is not 
changed. The generalization consists of feeding back the nonlinear 
mapping signal, rather than a signal linearized around the fixed 
point. This increases the basin of attraction of the 
controlled signal and thus decreases the sensitivity to noise. 
However, the range of parameters for which control can be achieved 
is limited. An addition to the control procedure, introduced 
by Socolar {\sl et. al}\cite{socolar}, is to allow memory of 
all previous periods. This latter procedure was implemented in the 
mapping parameter, rather than directly into the variables. The method 
allows an arbitrary range of the parameter in the logistic map to be 
tracked, but at the expense of a rapidly decreasing the basin of attraction 
with increasing range of parameter tracking. A generalization of the 
nonlinear feedback applied to the variables, to include memory, also 
allows arbitrary tracking of the parameter, with a significantly 
improved basin of attraction. All of these above control procedures 
increase the dimensionality of the phase space for a one-dimensional map 
to two. 

For control with memory there are two parameters to be chosen, the 
control parameter $K$ and the memory parameter $R$. For the general 
case, $R \not= K$, there is no superstable orbit. However, for 
nonlinear control with the choice $R=K$  the superstable orbit 
is recovered, and, remarkably, the phase space for the 
controlled logistic map is again one-dimensional. Because we recover 
the superstable orbit we call this an ``optimized" solution. 

The control methods we have been considering have in common with 
OGY the following properties: (a) the 
fixed points of the controlled map are the same as in the
uncontrolled system; (b) the feedback term vanishes in 
the absence of noise when control
is achieved; and (c)
one does not need to know the 
mapping equations in order
to apply the control. 
Unlike OGY, (d) no computer analysis of the system is necessary 
to apply the control and the methods probably can be applied for 
fast systems; and (e)  knowledge of the  location of the 
unstable periodic orbit 
is not necessary. For the ``optimized" control, (f) the 
dimensionality  of the controlled equations is the same
as in the uncontrolled system; 
(g) the control does not destroy the superstable orbit of the
uncontrolled system; 
while simultaneously, (h) 
control can be achieved in a very large region of the parameter space; 
(i) the basin of attraction of the controlled orbit is larger than 
in the other methods; and, consequently (j) the control 
is more  robust in the presence of  noise. However, (k) to achieve 
control with the parameter values that are deep within the unstable 
region, the nonlinear transient times to return to the controlled 
orbit becomes increasingly long.

Although we have only considered the application to mappings, of the 
various methods of control, the methods are also applicable to continuous 
systems governed by ordinary differential equations (ODE's). 
This was considered 
in the original paper by Pyragas\cite{pyragas}, who applied the linear 
control to the Roessler, Duffing and Lorenz systems. However, unlike a 
mapping, a simple delay makes the dimensionality of the system infinite. 
We can generalize our nonlinear feedback control for the case of 
ODE's. For the method without memory the $f(t)$ in 
Eq.~(\ref{eq1}) is replaced by 
$f(t)=-K[P({y}(t), {\bf x}(t)) - P({y}(t-\tau ),{\bf x}(t-\tau))]$, 
where  
$\tau $ is the period of the UPO.  
For the nonlinear control with memory Eq.~(\ref{eq1}) becomes 
${{dy(t)}\over{dt}}=P({y}(t),{\bf x} (t))+\epsilon(t)$, 
with
$\epsilon(t)=-K[P( y(t),{\bf x}(t)) - P({ y}(t-\tau ),{\bf x}(t-\tau))] +R \epsilon (t - \tau)$.  
In the case of the ``optimized" control we have for the $y$-equation  
${{dy(t)}\over{dt}}=(1-K)P({y}(t),{\bf x}(t))+K{{dy(t-\tau)}\over{dt}}$.  
We have achieved control of the Roessler system using all of these  
types of feedback. 
We are currently investigating which methods  
give the best performance with respect 
to the issues that we considered in this paper.

\acknowledgments
We thank M. A. Lieberman, J. Socolar and H-A. Tanaka for fruitful discussions. 
This work was partially supported by the
Office of Naval Research (Grant N00014-89-J-1097).

\begin{figure}
\caption[f2]{
Boundary of stability of the period-one orbit for the nonlinear control method 
without memory 
(solid line) and  Pyragas' linear control (dashed line). 
}
\label{f2}
\end{figure}  

\begin{figure}
\caption[f3]{
Liapunov exponent, $\log _2|\lambda |$, as a function of 
$a$ for the nonlinear control  
method without memory 
(solid line) with $K=1/3$ (the value of $K$ that gives the 
maximum $a^*$), Pyragas' linear control (dashed line) along 
the dotted line 
of Fig.~\ref{f2}, and  
for the uncontrolled logistic map (short-dashed line).  
}
\label{f3}
\end{figure}

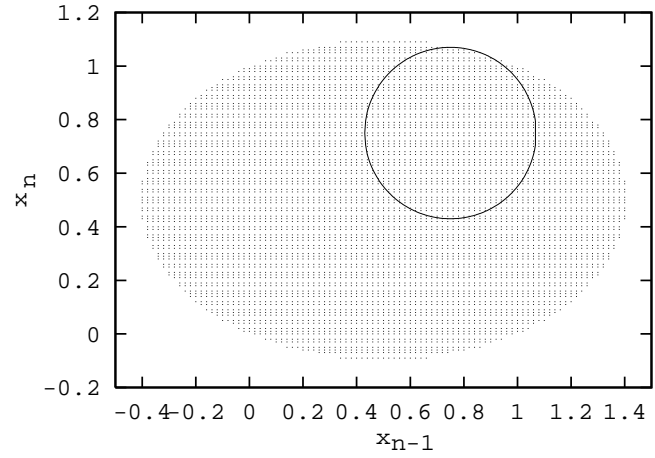
\begin{figure}
\caption[f5]{
Noise radius $r$  
as a function of $a$ for the nonlinear control method at $K=1/3$ (solid line) 
and for Pyragas' linear control along the dotted line of  Fig.~\ref{f2} 
(dashed line).  
}
\label{f5}
\end{figure}  

\begin{figure}
\caption[f7]{
Basin of attraction and noise circle for (a) our nonlinear control 
with memory at $K=0.6428$, 
and (b) for 
Socolar's control method at $K=1.8333$. In both cases, $R=0.5$  and $a=1$. 
These parameters  
correspond to the diamond symbol and 
cross shown in Fig.~\ref{f6}.  
}
\label{f7}
\end{figure}

\begin{figure}
\caption[f8]{
Rescaled noise radius $r$ 
as a function of $a$ for our nonlinear control with memory 
with $K=0.6428$ (that 
is, the value of $K$ that gives the maximum $a^*$ in  Fig.~\ref{f6}), and
Socolar's control method along the median line of the boundary of stability 
shown in  Fig.~\ref{f6}. In both cases  $R=0.5$.  
}
\label{f8}
\end{figure}

\begin{figure}
\caption[f12]{
Nonlinear transient for the logistic map along the line of the superstable 
orbit, $a_s$, shown in Fig.~\ref{f9}. 
}
\label{f12}
\end{figure}

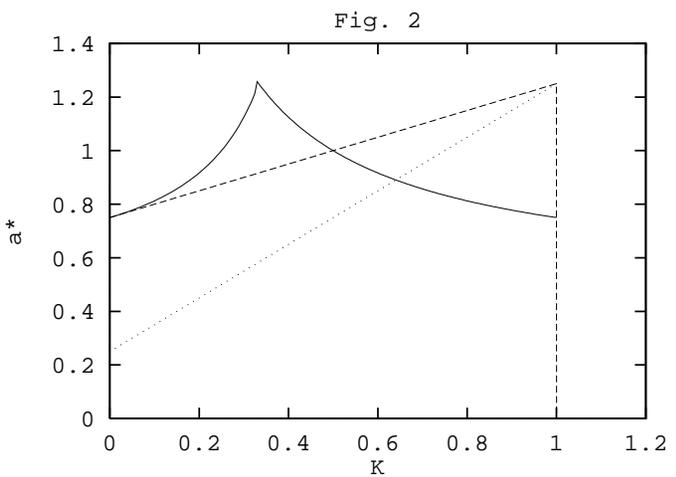
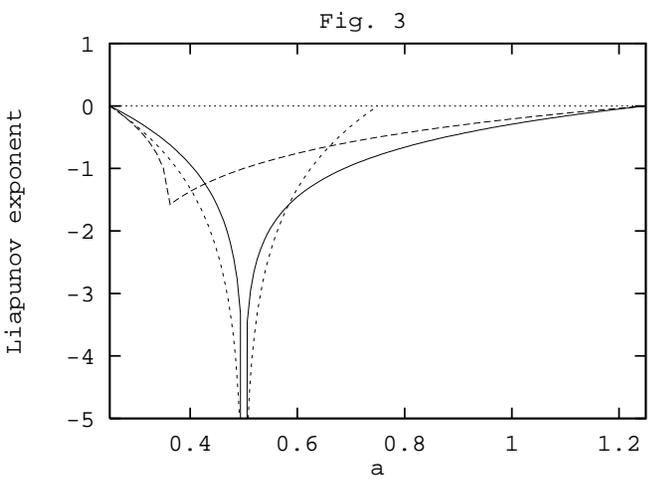
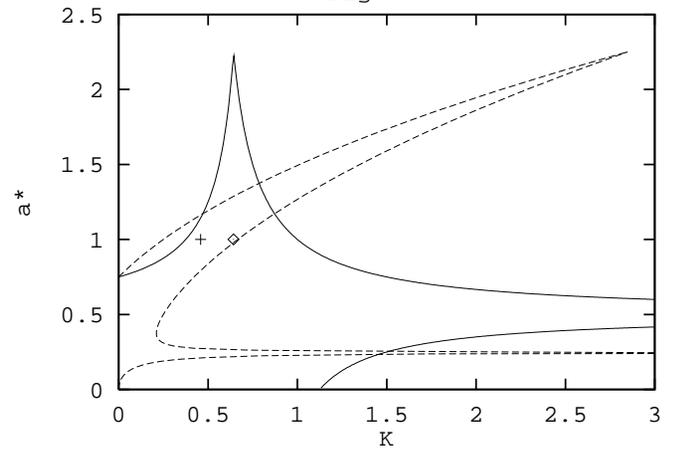
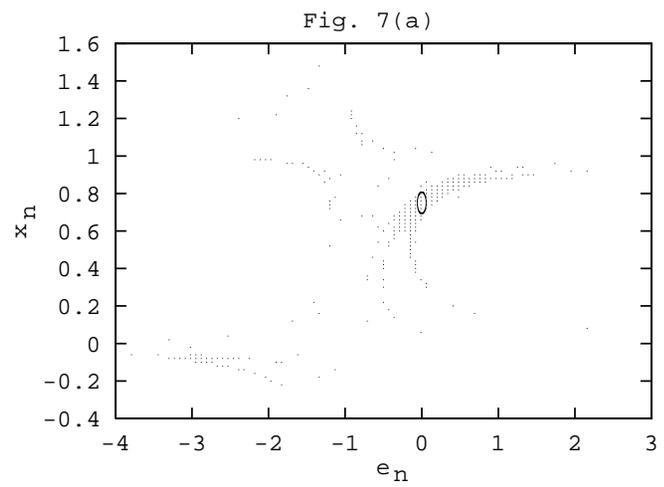
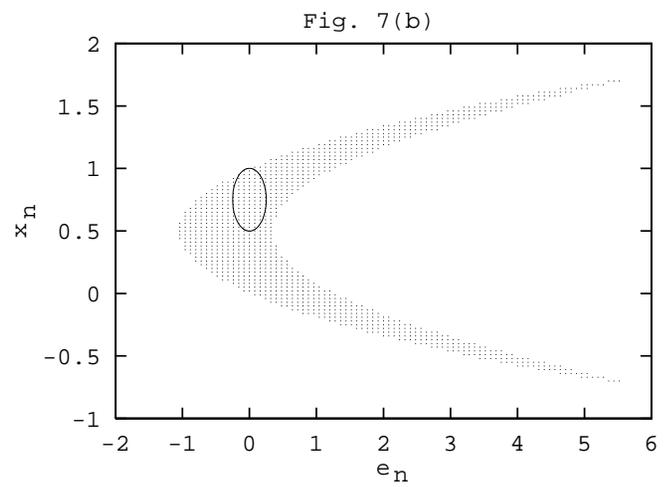
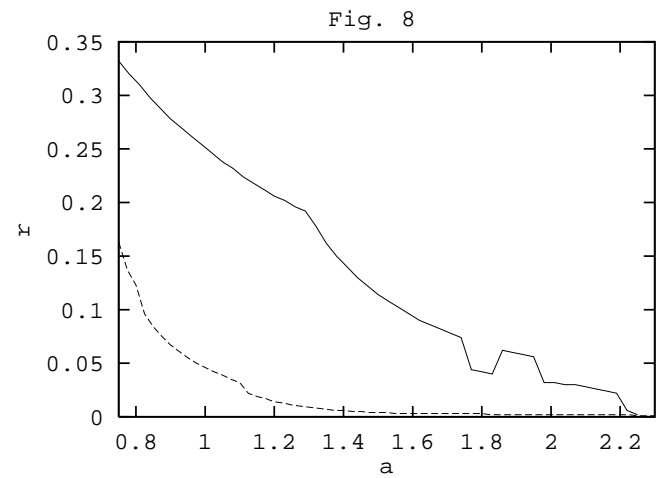
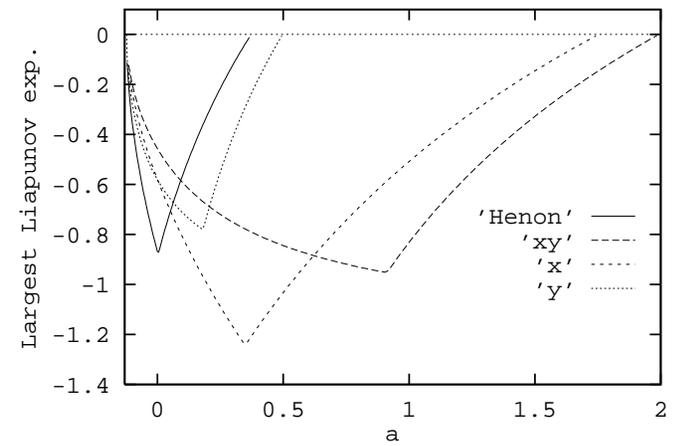
\end{document}